\documentstyle[11pt]{article}
\def\ft#1#2{{\scriptstyle {#1 \over #2}}}

\def\ww3{{$W_3$}}

\def\del{\partial}
\def\a{\alpha}

\def\ket#1{{| #1 \rangle}}

\begin{document}
\topmargin 0pt
\oddsidemargin 5mm
\begin{titlepage}
\begin{flushright}
CTP TAMU-2/95\\
hep-th/9504121\\
\end{flushright}
\vspace{1.5truecm}
\begin{center}
{\bf {\Large $N=1$ Superstring in $2+2$ Dimensions}}
\vspace{1.5truecm}

{\large Z. Khviengia, H. Lu\footnote{Supported in part by the
U.S. Department of Energy, under grant DE-FG05-91-ER40633},\ \  C.N.
Pope$^{1,}$, E. Sezgin\footnote{Supported in part by the National Science
Foundation, under grant PHY-9411543}, X.J. Wang and K.W. Xu}
\vspace{1.1truecm}

{\small Center for Theoretical Physics, Texas A\&M University,
                College Station, TX 77843-4242} \vspace{1.1truecm}


\end{center}
\vspace{1.0truecm}

\begin{abstract}
\vspace{1.0truecm}
In this paper we construct a $(2,2)$ dimensional string theory with manifest
$N=1$ spacetime supersymmetry.  We use Berkovits' approach of augmenting the
spacetime supercoordinates by the conjugate momenta for the fermionic
variables.   The worldsheet symmetry algebra is a twisted and truncated
``small'' $N=4$ superconformal algebra.  The realisation of the symmetry
algebra is reducible with an infinite order of reducibility.  We study the
physical states of the theory by two different methods.  In one of them, we
identify a subset of irreducible constraints, which is by itself critical.
We construct the BRST operator for the irreducible constraints, and study the
cohomology and interactions.  This method breaks the $SO(2,2)$ spacetime
symmetry of the original reducible theory.  In another approach, we study
the theory in a fully covariant manner, which involves the introduction of
infinitely many ghosts for ghosts.

\end{abstract}
\end{titlepage}
\newpage
\pagestyle{plain}
\section{Introduction}

     The Green-Schwarz superstring \cite{GS}, with manifest spacetime
supersymmetry, has  proved to be notoriously difficult to quantise in a
covariant manner.  The difficulty stems from the fact that there is no
kinetic term for the fermionic spacetime coordinates.   This problem has been
overcome recently by Berkovits \cite{Berk} in a reformulation of the
superstring, in which the spacetime supercoordinates are augmented by the
conjugate momenta for the fermionic variables.  The theory has $N=2$
worldsheet supersymmetry, as well as manifest four-dimensional $N=1$ spacetime
supersymmetry.   This theory can be thought of as a ten-dimensional theory
compactified on a Calabi-Yau background.

     It is interesting to investigate whether such an approach could be used
for constructing an intrinsically four-dimensional theory with manifest
spacetime supersymmetry.  This would contrast strikingly with the
$N=2$ NSR string \cite{OV} which, although it has a four-dimensional
spacetime (with $(2,2)$ signature), has no supersymmetry in spacetime.  In
fact it has only one Neveu-Schwarz state, describing self-dual Yang-Mills in
the open string, and self-dual gravity in the  closed string \cite{OV},
together with an additional Ramond massless state which is also bosonic
\cite{lp}.  Attempts have been made to find a supersymmetric version of the
theory.  In a recent paper \cite{BKL}, it was  observed that massless
fermionic physical states, as well as bosonic ones, appear in certain $Z_2$
twisted sectors of the theory.  This is however at the price of breaking the
spacetime structure, and in addition it does not have the usual definition
of spacetime supersymmetry.

    If a four-dimensional string of the Berkovits type could be constructed,
it would be quite different from the above case, in that it would have a
manifest spacetime supersymmetry.  A way to build such a theory is suggested
by some work of  Siegel \cite{S1}.  He considered a set  of quadratic
constraints built from the coordinates and momenta of a  superspace in $2+2$
dimensions, and thus displaying manifest spacetime  supersymmetry.  In
\cite{S1} it was proposed that in the case of open  strings, this theory
described self-dual $N=4$ super Yang-Mills, whilst the  corresponding closed
string described self-dual $N=8$ supergravity.

     The full set of constraints considered in \cite{S1} do not generate a
closed algebra.   However, we find that there exists a subset of the
constraints that does close on an algebra, with two bosonic spin-2
generators and two fermionic spin-2 generators.  In this paper we build a
Berkovits-type open string theory in four dimensions, based on this
worldsheet symmetry algebra.  Noting that the central charge in the ghost
sector vanishes, we see that the matter fields should also have zero central
charge.  We achieve this by taking the coordinates $(X^\mu,\theta^\alpha)$
of a chiral $N=1$ superspace, together with the canonical momenta $p_\alpha$
for the fermionic coordinates.  This is a chiral restriction of the
analogous matter system introduced by Berkovits \cite{Berk}.

     The chiral truncation that we are making is possible only if the
signature of the four-dimensional spacetime is $(2,2)$.  In this case, the
$SO(2,2)$ Lorentz group is the direct product $SL(2,R)_{\rm L} \times
SL(2,R)_{\rm R}$, with dotted spinorial indices transforming under
$SL(2,R)_{\rm L}$, and undotted indices transforming under $SL(2,R)_{\rm R}$.
The bosonic currents are singlets under the entire Lorentz group,
but the two fermionic currents form a doublet under $SL(2,R)_{\rm L}$.

     Unfortunately, these constraints are reducible, which implies that the
associated ghosts still have gauge invariances, whose elimination requires
the introduction of ghosts for ghosts.  In fact the reducibility is of
infinite order, just as in the covariant Green-Schwarz superstring, and thus
an infinite number of ghosts for ghosts are needed.   This problem can be
overcome at the price of sacrificing manifest spacetime Lorentz invariance,
since in this case we can identify a subset of the constraints that is
irreducible, and which also has the critical central charge.  Essentially,
the states that are physical under this irreducible subset are also
annihilated by the the remaining dependent constraints. (This statement will
be made precise for states with standard ghost structure.)   In section 2,
we discuss the full set of constraints and their irreducible subset. Since
the irreducible system is also critical, we obtain the corresponding BRST
operator.  The irreducible system still maintains $N=1$ spacetime
supersymmetry.  In section 3, we construct some examples of physical states
of the BRST operator for the irreducible system, and discuss their
interactions.  There are two massless physical states which have standard
ghost structure, namely a scalar and its spin-$\ft12$ superpartner.  The
physical spectrum also contains an infinite number of massive states.

     Although the irreducible system can be solved completely, the spacetime
$SO(2,2)$ covariance of the original reducible system is broken.  It is of
interest to have a fully covariant BRST treatment.   This system is in
one respect slightly simpler than previous examples of reducible systems,
such as the covariant Green-Schwarz superstring and the $N=4$ string, in
that one can build a nilpotent charge $Q'$ from the reducible constraints.
This implies that the standard ghost vacuum of the ghosts for ghosts has
zero conformal dimension.  Thus if we restrict our attention to physical
states that are the tensor product of this vacuum with states in the
cohomology of $Q'$, the physical-state condition simplifies
considerably, and can be discussed without the need to know all the
details of the fully covariant BRST operator.  We shall discuss the fully
covariant BRST procedure for the reducible system in section 4, and discuss
the corresponding cohomology in section 5.   We shall see that indeed that
extra conditions arising from ghosts for ghosts eliminate states of $Q'$
that have the undesirable feature of carrying infinite-dimensional
$SL(2,R)_{\rm L}$ representations.   The remaining states of the reducible
system, after a further truncation of non-interactive states, are expected
to coincide with those of the irreducible system.  In particular we show
that the massless physical states give rise to the identical interactions.

\section{The constraint algebra and the BRST charge}

     In this section, we discuss the algebra of constraints that defines the
string theory.  The matter system consists of the four spacetime coordinates
$X^{\alpha\dot\alpha} = \sigma^{\alpha\dot\alpha}_\mu\, X^\mu$, the
two-component Majorana-Weyl spinor $\theta^\alpha$, and its conjugate
momentum $p_\alpha$.  The action for the matter system takes the form
\begin{equation}
I=\int d^2z\, \Big( -\ft12 \del X^{\alpha\dot\alpha}\, \bar\del X_{\alpha
\dot\alpha} + p_\alpha\, \bar\del \theta^\alpha \Big) \ .
\end{equation}
In the language of conformal field theory, these fields satisfy the OPEs
\begin{equation}
X^{\alpha\dot\alpha}(z) X^{\beta\dot\beta}(w) \sim - \epsilon^{\alpha\beta}
\, \epsilon^{\dot\alpha\dot\beta} \log(z-w),\qquad \qquad p_\alpha(z)
\theta_{\beta}(w) \sim {\epsilon_{\alpha\beta}\over z-w} \ .
\label{matopes}
\end{equation}
When we need to be explicit, we use conventions in which the spacetime metric
is given by
$\eta_{\mu\nu}={\rm diag}\, (-1,-1,1,1)$, the indices $\mu,\nu\,\ldots$ run
from 1 to 4, and the mapping between tensor indices and 2-component spinor
indices is defined by
\begin{equation}
V^{\alpha\dot\alpha}=
\pmatrix{V^{1\dot1} & V^{1\dot2}\cr
         V^{2\dot1} & V^{2\dot2}\cr} = {1\over\sqrt 2}\,
\pmatrix{V^1+V^4 & V^2-V^3 \cr
         V^2+V^3 &-V^1+V^4\cr}\ ,
\label{vdw}
\end{equation}
where $V^\mu$ is an arbitrary vector.  The Van der Waerden symbols
$\sigma^{\alpha\dot\alpha}_\mu$ thus defined satisfy
$\sigma^{\alpha\dot\alpha}_\mu\, \sigma^{\mu\, \beta\dot\beta} =
\epsilon^{\alpha\beta}\, \epsilon^{\dot\alpha\dot\beta}$, where
$\epsilon_{12}=\epsilon^{12}=1$.  Spinor indices are
raised and lowered according to the usual ``North-west/South-east''
convention, with $\psi^\alpha= \epsilon^{\alpha\beta}\, \psi_\beta$ and
$\psi_\alpha=\psi^\beta\, \epsilon_{\beta\alpha}$, {\it etc.}, so we have
$\psi^1=\psi_2$ and $\psi^2=-\psi_1$.  Note that since the indices are
two-dimensional, we have the useful Schoutens identity $X^\alpha\,
Y_\alpha\, Z_\beta + X_\alpha\, Y_\beta\, Z^\alpha + X_\beta\,
Y^\alpha\, Z_\alpha =0$.

     In \cite{S1}, Siegel proposed to build a string theory implementing the
set of constraints given by $\Big\{\del X^{\alpha\dot\alpha}\del
X_{\alpha\dot\alpha},\, p_\alpha\,\del\theta^\alpha, \,p_\alpha p^\alpha,\,
\del\theta_\alpha\, \del\theta^\alpha,\, p_\alpha\,\del
X^{\alpha\dot\alpha},\,
\del\theta_\alpha\, \del X^{\alpha\dot\alpha}\Big\}$.  However, it follows
from (\ref{matopes}) that whilst the second order poles in the OPEs
amongst this set of constraints give back the same set of constraints, not
all the first-order poles can be re-expressed as the derivatives of the
constraints.  In other words, the algebra does not close.  (Note that this
non-closure occurs even at the classical level of Poisson brackets, or
single OPE contractions.)  Accordingly, we
choose a subset of Siegel's constraints that form a closed algebra, namely
\begin{eqnarray}
T &=& -\ft12 \del X^{\alpha\dot\alpha}\, \del X_{\alpha\dot\alpha} - p_\alpha\,
\del\theta^\alpha \ ,\nonumber \\
\label{matcur}
S &=& - p_\alpha\, p^\alpha\ ,\\
G^{\dot\alpha} &=& - p_\alpha\, \del X^{\alpha\dot\alpha}\ .\nonumber
\end{eqnarray}
We see from the energy-momentum tensor that from the worldsheet viewpoint,
$\theta^\alpha$ has conformal weight 0, and $p_\alpha$ has conformal weight
1.  Thus the two bosonic currents $T$ and $S$, and also the two fermionic
currents $G^{\dot\alpha}$, have conformal spin 2.  The currents are all
primary, and the remaining non-trivial OPE is given by
\begin{equation}
G^{\dot\alpha}(z) G^{\dot\beta}(w) \sim {2\epsilon^{\dot\alpha\dot\beta} \,
S\over (z-w)^2} + {\epsilon^{\dot\alpha\dot\beta}\,\del S\over z-w} \ .
\end{equation}

     The matter currents may be expressed in a concise form by introducing a
pair of spin-0 fermionic coordinates $\zeta^{\dot\a}$ on the worldsheet.  We
can then define
\begin{equation}
{\cal P}^\a= p^\a +\zeta_{\dot\a}\, \del X^{\a\dot\a} + \zeta_{\dot\a}\,
\zeta^{\dot\a}\, \del\theta^\alpha\ ,\label{psuper}
\end{equation}
in terms of which the currents may be written as ${\cal T}={\cal P}_\a\,
{\cal P}^\a$, where
\begin{equation}
{\cal T}= S + \zeta_{\dot\a}\, G^{\dot\a} + \zeta_{\dot\a}\,
\zeta^{\dot\a}\, T\ .\label{tsuper}
\end{equation}

    Note that the algebra generated by (\ref{matcur}) is a truncation of the
``small'' $N=4$ superconformal algebra that was used to construct the $N=4$
string in \cite{S2}.   This can be seen from the fact that we can augment
our currents (\ref{matcur}) by including $\Big\{\theta_\alpha
\theta^\alpha,\ p_\alpha\, \theta^\alpha,\, \theta_\alpha\, \del
X^{\alpha\dot\alpha}\Big\}$ as well.  One can easily verify that the
resulting currents generate precisely the small $N=4$ superconformal
algebra, in a twisted basis.  It was shown in \cite{S2} that this
realisation of the $N=4$ algebra is reducible.  In fact the constraints of
its $N=2$ subalgebra are irreducible, and the associated physical states are
also annihilated by the full $N=4$ currents.  Thus in $2+2$ dimensions,
the $N=4$ string is equivalent to the $N=2$ string \cite{S2}, which is
generally believed not to have spacetime supersymmetry.   Our choice of
currents (\ref{matcur}), which is motivated by the desire to obtain a string
theory in $2+2$ dimensions which does have spacetime supersymmetry, generates
a different subalgebra of the $N=4$ algebra.

    Unfortunately, the currents given in (\ref{matcur}) are also reducible.
Specifically, one can observe that
\begin{equation}
p^\a\, T + \del X^{\a\dot\a}\, G_{\dot\a} + \del\theta^\a\, S=0
\ ,\qquad
p^\a\, G^{\dot\a} + \del X^{\a\dot\a}\, S = 0\ , \qquad p_\a\, S=0\ .
\label{reduceq}
\end{equation}
These can be written in the concise form ${\cal P}_\a\, {\cal T}=0$.
As in the case of \cite{S2}, the reducible constraints (\ref{matcur}) can be
divided into independent constraints and dependent constraints.  The
independent constraints can be taken to be
\begin{equation}
T = -\ft12 \del X^{\alpha\dot\alpha}\, \del X_{\alpha\dot\alpha} - p_\alpha\,
\del\theta^\alpha\ , \qquad
G^{\dot1} = - p_\alpha\, \del X^{\alpha\dot1}\ ,\label{n1matcur}
\end{equation}
which in fact generate a subalgebra of the twisted $N=2$ superconformal
algebra. Using (\ref{reduceq}), we can write the remaining constraints, {\it
i.e.~}the dependent ones, as linear functions of the independent
constraints.  In momentum space, they are given by
\begin{equation}
S = (p^{\a\dot1})^{-1}\, p^\a\, G^{\dot1} \ ,\qquad
G^{\dot2} =- (p^{\a\dot1})^{-1}
(p^\a\, T + p^{\a\dot2}\, G^{\dot1} +\del\theta^\a\, S)\ ,\label{dependency}
\end{equation}
where $\a$ can be chosen to be either 1 or 2.   These expressions are valid
in the region of phase space where $p^{\a\dot1} \ne 0$.   To cover the
region where $p^{\a\dot1} = 0$, we can make a different choice of the
independent constraints.

      The above reducibility of the constraints can be better understood by
studying the physical spectrum of the theory.  First let us consider the
physical operators with standard ghost structure, in which case knowledge of
the explicit form of the BRST operators is not necessary.  As we shall see
later, after imposing the $T$ and $G^{\dot1}$ constraints, there are
two massless physical operators with standard ghost structure, which
form a spacetime $N=1$ supermultiplet.  This pair of operators is then
identically annihilated by the remaining constraints $G^{\dot2}$ and $S$.
This establishes the equivalence between the constraints of (\ref{matcur})
and the reduced ones (\ref{n1matcur}), for the massless physical states.
However there are further massive operators with standard ghost structure
under only the $T$ and $G^{\dot1}$ constraints, which do not seem to be
annihilated by the constraints of $G^{\dot2}$ and $S$.  To establish the
equivalence of the massive spectra of the reducible and the irreducible
systems would require the analysis of the full cohomology and interactions,
including the physical states with non-standard ghost structure.

      In order to discuss the physical spectrum with non-standard ghost
structure, it is necessary to obtain the explicit form of the BRST
operators.    The construction of the BRST operator for a system with
reducible constraints is discussed in \cite{fhst}.   The reducibility
implies that the ghosts for the original constraints still have gauge
invariances, whose elimination requires the introduction of ghosts for
ghosts.   As in the case of covariant quantisation of the Green-Schwarz
string, the reducibility relations (\ref{reduceq}) are themselves
overcomplete; in fact the system has an infinite order of reducibility.
This can be easily seen from the form ${\cal P}_\a\, {\cal T}=0$ for the
reducibility relations, owing to the fact that the functions ${\cal P}_\a$
are themselves reducible, since ${\cal P}_\a\, {\cal P}^\a$ gives back the
constraints ${\cal T}$.  This infinite order of reducibility implies that a
proper BRST treatment requires an infinite number of ghosts for ghosts.  The
form of the BRST operator, after making the decomposition into independent
constraints $(T, G^{\dot1})$ and dependent constraints $(G^{\dot2}, S)$, is
\begin{equation}
{\widetilde Q} = Q + \sum_{k\ge 0} \hat b_{\a_k}\, c^{\a_{k+1}}\ ,
\end{equation}
where $Q$ is the standard BRST operator for the irreducible system described
by $(T, G^{\dot1})$, $\hat b_{\a_{k}}$ are the level-$k$ antighosts for the
dependent constraints, and $c^{\a_{k}}$ are the level-$k$ ghosts for the
independent constraints.   Since the ghost fields in the second term do not
appear in $Q$, and they form Kugo-Ojima quartets, the BRST cohomology of
$\widetilde Q$ is equivalent to that of $Q$.   To see this, note that any
states with excitations of $\hat c^{\a_k}$ or $b_{\a_k}$ will not be
annihilated by the BRST operator, whilst any states with excitations of
$\hat b_{\a_k}$ or $c^{\a_k}$ are BRST trivial, since $\{\widetilde Q,
b_{\a_{k+1}}\} =\hat b_{\a_k}$ and $\{\widetilde Q, \hat c^{\a_{k-1}}\} =
-c^{\a_k} $.  Although the BRST operator $\widetilde Q$ was obtained at the
classical level, it is also nilpotent at the quantum level.

     The price that we have paid for the simple form of the BRST operator
$\widetilde Q$ is that the $SL(2,R)_{\rm L}$ covariance of the $\dot\a$
indices in the original reducible constraints has been sacrificed.   The
fully covariant BRST treatment of the reducible system remains to be
understood.  We shall now proceed by constructing the BRST operator $Q$ for
the irreducible system $(T, G^{\dot1})$.   We begin by introducing the
anticommuting ghosts $(b,c)$ and the commuting ghosts $(r,s)$ for $T$ and
$G^{\dot1}$ respectively.  The commuting ghosts $(r, s)$ are bosonised, {\it
i.e.} $r=\del\xi\, e^{-\phi}$, $s=\eta\, e^{\phi}$. In terms of these
fields, the BRST operator $Q$ is given by
\begin{eqnarray}
Q&=& c \Big(-\ft12 \del X^{\a\dot\a}\, \del X_{\a\dot\a} - p_\a\,
\del\theta^\a - b\, \del c -\ft12 (\del\phi)^2 -\ft32 \del^2\phi - \eta\,
\del \xi\Big) \nonumber\\
&+& \eta\, e^{\phi}\, p_\a\,\del X^{\a\dot1}\ .\label{n1brst}
\end{eqnarray}

      The theory has spacetime supersymmetry, generated by
\begin{eqnarray}
q^\a &=& \oint p^\a\ ,\nonumber\\
q^{\dot1} &=&\oint \theta_\a\, \del X^{\a\dot1}\ ,\qquad
q^{\dot2} =\oint \theta_\a\, \del X^{\a\dot2} + b\,\eta\,e^{\phi}\ .
\label{n1susygen}
\end{eqnarray}
The somewhat unusual ghost terms in $q^{\dot2}$ are necessary for the
generator to anti-commute with the BRST operator. It is straightforward to
verify that these supercharges generate the usual $N=1$ spacetime
superalgebra
\begin{equation}
\{q_\alpha,q_\beta\}=0= \{q^{\dot\alpha},q^{\dot\beta} \},\qquad
\{q^\alpha,q^{\dot\alpha} \}= P^{\alpha\dot\alpha} \ ,\label{n1susyalg}
\end{equation}
where $P^{\alpha\dot\alpha}=\oint \del X^{\alpha\dot\alpha}$ is the
spacetime translation operator.

     Since the zero mode of $\xi$ is not included in the Hilbert space of
physical states, there exists a BRST non-trivial picture-changing operator
$Z=\{Q, \xi\}$ which can give new BRST non-trivial physical operators when
normal ordered with others.  Explicitly, it takes the form
\begin{equation}
Z=c\,\del \xi + p_\a\, \del X^{\a\dot1} e^{\phi}\ . \label{n1pic}
\end{equation}
Unlike the picture-changing operator in the usual $N=1$ NSR superstring,
this operator has no inverse.

\section{Physical states and interactions}

    In this section, we shall discuss the cohomology of the BRST operator
$Q$ given in (\ref{n1brst}) for the irreducible system $(T, G^{\dot1})$, and
present some results for the physical states in the theory.  We begin by
studying the physical spectrum with standard ghost structure.  There are two
massless operators
\begin{equation}
V=c\, e^{-\phi}\, e^{ip\cdot X}\ ,\qquad
\Psi = h_\a\, c\, e^{-\phi}\, \theta^\a\, e^{ip\cdot X}\ ,\label{n1mass0}
\end{equation}
which are physical provided with mass-shell condition $p^{\a\dot\a}\,
p_{\a\dot\a} = 0$ and spinor polarisation condition $p^{\a\dot1}\, h_a =0$.
The non-triviality of these operators can be established by the fact that
the conjugates of these operators with respect to the following
non-vanishing inner product
\begin{equation}
\Big\langle \del^2c\,\del c\, c\, e^{-3\phi}\, \theta^2 \Big\rangle
\label{n1innpro}
\end{equation}
are also annihilated by the BRST operator. The bosonic operator $V$ and the
fermionic operator $\Psi$ form a supermultiplet under the $N=1$ spacetime
supersymmetric transformation.  The associated spacetime fields $\phi$ and
$\psi_\a$ transform as
\begin{equation}
\delta\phi = \epsilon_\a\, \psi^\a\, \qquad
\delta\psi_\a = \epsilon^{\dot\a}\, \del_{\a\dot\a}\, \phi\ .
\end{equation}

      We can build only one three-point amplitude among the massless
operators, namely
\begin{equation}
\Big\langle V(z_1)\,\, \Psi(z_2)\,\, \Psi(z_3)\Big\rangle =c_{23}\ ,
\label{n13pf}
\end{equation}
where $b_{ij}$ is defined by
\begin{equation}
b_{ij} =  h_{(i)\a}\,h^\a_{(j)}\ .\label{n13pf1}
\end{equation}
{}From this, we can deduce that the $V$ operator describes a spacetime scalar
whilst the $\Psi$ operator describes a spacetime chiral spin-$\ft12$
fermion.  Note that this is quite different from the case of the $N=2$
string where there is only a massless boson and although it is ostensibly a
scalar, it in fact, as emerges from the study of the three-point amplitudes,
a prepotential for self-dual Yang-Mills or gravity.  With the one insertion
of the picture-changing operator, we can build a four-point function which
vanishes for kinematic reasons:
\begin{equation}
\Big\langle ZV\, \Psi\, \oint b\Psi\, \Psi \Big\rangle  =
(u\, b_{12}\, b_{34} + s\, b_{13}\, b_{24}){\Gamma(-\ft12 s)
\, \Gamma(-\ft12 t)\over \Gamma(\ft12 u)}\ ,\label{4pfmass0}
\end{equation}
where $s$, $t$, and $u$ are the Mandelstam variables and $h_{(1)}^\a =
p^{\a\dot1}_{(1)}$.  The vanishing of
the kinematic term, {\it i.e.}
\begin{equation}
u\, b_{12}\, b_{34} + s\, b_{13}\, b_{24} =0\label{identity1}
\end{equation}
is a straightforward consequence of the mass-shell condition of the
operators and momentum conservation of the four-point amplitude \cite{lp}.
It might seem that the vanishing of the this four-point amplitude should be
automatically implied by the statistics of the operators since there is an
odd number of fermions.   However, as we shall see later, the
picture-changing operator has spacetime fermionic statistics.   In fact,
that the four-point amplitude (\ref{4pfmass0}) vanishes only on-shell, for
kinematic reasons (\ref{identity1}),  already implies that the picture
changer $Z$ is a fermion. Thus the picture changing of a physical operator
changes its spacetime statistics and hence does not establish the
equivalence between the two.  On the other hand, since $Z^2=(ZZ)$ becomes a
spacetime bosonic operator,  we can use $Z^2$ to identify the physical
states with different pictures.  Thus we have a total of four massless
operators, namely $V$, $ZV$ and their supersymmetric partners.  $V$ and its
superpartner $\Psi$ have standard ghost structure; $ZV$ and its superpartner
$Z\Psi$ have non-standard ghost structures.

      So far we have constructed massless physical states.  There are also
infinitely many massive states.   The tachyonic type of massive
operators, {\it i.e.}~those that become pure exponentials after bosonising
the fermionic fields, can be easily obtained.  They are given by
\begin{eqnarray}
V_n&=& c(\del^np)^2\cdots p^2\, e^{n\phi}\, e^{ip\cdot X}\ ,\qquad
{\cal M}^2 =(n+1)(n+2)\ ,  \nonumber\\
\widetilde V_n &=& c(\del^{n+1}\theta)^2\cdots \theta^2\, e^{-(n+3)\phi}
\, e^{ip\cdot X}\ ,\qquad
{\cal M}^2 = (n+1)(n+2)\ ,\nonumber\\
U^{(\delta)}_n&=& h^{(\delta)}_\a\, c\,
\del^n p^\a\, (\del^{n-1} p)^2\cdots p^2\,
e^{(n-1+\delta)\phi}\, e^{ip\cdot X}\ ,\qquad {\cal M}^2 =
(n+1)(n+2-2\delta)\ ,\nonumber\\
\widetilde U^{(\delta)}_n &=& \tilde h^{(\delta)}_\a\, c\,
\del^{n+1}\theta^\a\, (\del^{n}\theta)^2
\cdots \theta^2\, e^{-(n+2+\delta)\phi}\, e^{ip\cdot X}\ ,
\qquad {\cal M}^2 = (n+1)(n+2-2\delta)\ ,\label{n1masstats}\\
W_n &=& h_{\a\beta}\, c\, \del^{n+1} p^\a\, \del^n p^{\beta}
(\del^{n-1} p)^2\cdots p^2\, e^{n\phi}\, e^{ip\cdot X}\ ,
\qquad {\cal M}^2 = (n^2 + 3n +4)\ ,\nonumber\\
\widetilde W_n &=& \tilde h_{\a\beta}\, c\, \del^{n+2}\theta^\a\,
\del^{n+1}\theta^\beta\, (\del^n\theta)^2\cdots \theta^2\,
e^{-(n+3)\phi}\, e^{ip\cdot X}\ ,
\qquad {\cal M}^2 = (n^2 + 3n +4)\ ,\nonumber
\end{eqnarray}
where $p^2 = p_\a\, p^\a$, {\it etc.~}and $\delta=0$ or 1.  $V_n$ and
$\widetilde V_n$ are physical provided the proper mass-shell condition is
satisfied.  For the remaining operators, in addition to mass-shell
conditions, they also satisfy certain spinor polarisation conditions:
$p^{\a\dot1}\, h^{(1)}_\a = 0$, $p^{\a\dot1} \tilde h^{(0)}_\a =0$,
$p^{\a\dot2}\, h^{(0)}_\a = 0$, $p^{\a\dot2} \tilde h^{(1)}_\a =0$,
$p^{\a\dot1}\, h_{\a\beta} = 0$ and $ p^{\a\dot1}\, \tilde h_{\a\beta}=0$.
In addition $h_{\a\beta}$ and $\tilde h_{\a\beta}$ must be symmetric.

     The spacetime statistics of the physical operators can be determined as
follows. The non-vanishing of the three-point amplitudes $\langle \widetilde
V_{2n}\, U^{(1)}_n\, U^{(1)}_n\rangle$ and $\langle V_{2n+1}\, \widetilde
U^{(0)}_n\,\widetilde U^{(0)}_n \rangle$ implies that $V_{2n+1}$ and
$\widetilde V_{2n}$ are bosons. Since $\del c\,\widetilde V_{2n+1}$ and
$\del c\, V_{2n}$ are their conjugates, it follows that all $V_{n}$ and
$\widetilde V_n$ are bosons. On the other hand, the non-vanishing of the
three-point amplitudes $\langle \Psi\, U^{(1)}_n\, \widetilde V_{n-1}
\rangle$ and $\langle \Psi\, \widetilde U^{(0)}_n\, V_n\rangle$ leads to the
conclusion that $U^{(\delta)}_n$ and $\widetilde U^{(\delta)}_n$ are
fermions. There are various relations among the above physical operators.
For example, $Z U^{(0)}_n = V_n$ and $ZV_n = U^{(1)}_{n+1}$.   Thus we can
immediately see that the picture changer $Z$ is a spacetime fermion, and
hence the above relations do not establish the equivalence between the $U$
and $V$ operators.   However, they do imply that $U^{(0)}_n$ and $U^1_{n+1}$
are equivalent.   The superpartners of these operators can be obtained by
the action of the supersymmetry generators given in (\ref{n1susygen}).

      The above operators are only a small subset of the complete
spectrum of massive operators.  The ${(\rm mass)}^2$ of these operators grows
quadratically with $n$.  However, as we shall show, one can build
non-vanishing four-point amplitudes from these operators, which implies the
existence of massive operators with the usual linear growth.
The simplest four-point amplitude that can be built is given by
\begin{equation}
\Big\langle U^{(1)}_3\, \widetilde U^{(0)}_0\, \oint b
\widetilde U^{(0)}_0\,
\widetilde U^{(0)}_0\Big\rangle = \Big( A(\ft12 u-3) + B (\ft12 s -3)\Big)
{\Gamma(-\ft12s + 3)\Gamma(-\ft12 t + 3) \over \Gamma(\ft12 u - 2)}\ ,
\label{n1mass4pf}
\end{equation}
where $s, t, u$ are the Mandelstam variables, and $A, B$ are given by
\begin{equation}
A=h^\a_{(1)}\, h_{(2)\a}\, h_{(3)\beta}\, h^\beta_{(4)}\ ,\qquad
B=h^\a_{(1)}\, h_{(2)\beta}\, h_{(3)\a}\, h^\beta_{(4)}\ .
\end{equation}
The physical-state conditions imply that $p_{(1)}^{\a\dot\a}\,
p_{(1)\a\dot\a} = -12$, $p_{(i)}^{\a\dot\a}\, p_{(i)\a\dot\a} = -2$, for
$i=2,3,4$, and $p_{(i)}^{\a\dot1}\, h_{(i)\a} = 0$ for $i=1,2,3,4$.  Thus
the solution for these polarisation spinors is $h_{(i)}^{\a} =
p_{(i)}^{\a\dot1}$ (up to arbitrary scaling factors).   One can then easily
verify that the prefactor of the four-point amplitude (\ref{n1mass4pf}) does
not vanish.   This implies that there are further massive operators with
${(\rm mass})^2 = 2(n+3)$ in the spectrum.  Moreover this result is also
consistent with the fact that the picture-changing operator $Z$ is a
spacetime fermion.  To see this, we first note that the four-point amplitude
({\ref{n1mass4pf}) can be restated as $\langle ZV_2\, \widetilde U^{(0)}_0\,
\widetilde U^{(0)}_0\, \widetilde U^{(0)}_0\rangle$.  Since $V_2$ is
spacetime boson whilst $\widetilde U^0_0$ is spacetime fermion, it follows
that the non-vanishing of the four-point amplitude implies that the picture
changer $Z$ is a spacetime fermion.

      The massive physical operators that we found explicitly in
(\ref{n1masstats}) all have non-standard ghost structures.  From these
operators, we can build non-vanishing four-point amplitudes, which
implies the existence of further massive operators in the physical spectrum.
The structure of these massive operators that are exchanged in the
four-point amplitudes can be determined by the structures of the external
physical operators.  In particular, the non-vanishing four-point amplitude
$\langle U^{(1)}_n\, \widetilde U^{(0)}_{n-1}\, \Psi\, \Psi\rangle$ implies,
by looking at the $s$ channel, the existence of massive operators with
standard ghost structure.\footnote{We thank the referee for drawing our
attention to the existence of massive operators with standard ghost
structure.}

\section{Covariant quantisation of the reducible system}

       In the previous section, we discussed the physical spectrum and
interactions for the irreducible system $(T, G^{\dot1})$.   Although the
theory is spacetime supersymmetric, the manifest $SO(2,2)\sim SL(2,R)_{\rm L}
\times SL(2,R)_{\rm R}$ covariance of the original reducible system is
partially broken down to $SL(2,R)_{\rm R}$.  The fully covariant BRST
treatment of the original reducible system is far more complicated, since
the ghost for ghost terms no longer decouple.  This is a consequence of
the fact that the reducibility relations ${\cal P}^\a\, {\cal T} = 0$
have an infinite order of
reducibility since ${\cal P}^\a\, {\cal P}_\a = {\cal T}$,
which is reminiscent of the situation for the covariant quantisation of
the Green-Schwarz superstring.   Another example is the string theory in
$2+2$ dimensions associated with the small $N=4$ superconformal algebra,
which was discussed by Siegel in \cite{S2}.  Naively
the central charge for the $N=4$ string is $-12$; however, in $2+2$
dimensions the constraints are reducible.  The irreducible subset of the
constraints generates the $N=2$ superconformal algebra, for which the
critical dimension is indeed four.  Thus the proper fully covariant BRST
treatment of the reducible $N=4$ string requires the introduction of ghosts
for ghosts that will contribute to the central charge for criticality.  In one
respect our example is slightly simpler, in that the matter has the
critical central charge both for the irreducible system and the original
reducible system.   This means that we can write down a nilpotent charge $Q'$
for the reducible system without the need of ghost for ghost terms.   It
is of course not the true BRST charge for the corresponding string theory.
However since $Q'$ is already nilpotent, the contributions to the central
charge from the ghosts for ghosts should be zero.  We shall see later that
this feature makes the analysis of the cohomology much simpler.

      The fully covariant BRST procedure for a reducible system is described
in \cite{fhst}.   As we showed in section 2, the matter currents
can be expressed in a concise form (\ref{tsuper}) and (\ref{psuper}) by
introducing a pair of spin-0 fermionic coordinates $\zeta^{\dot\a}$ on the
worldsheet.  Analogously, the ghosts and antighosts can also be written in
the form:
\begin{eqnarray}
{\cal C}_k&=& c_k + \zeta_{\dot\a}\, s^{\dot\a}_k + \zeta_{\dot\a}
\zeta^{\dot\a}\, \gamma_k\ ,\nonumber\\
{\cal B}_k&=& \beta_k + \zeta_{\dot\a}\, r^{\dot\a}_k + \zeta_{\dot\a}
\zeta^{\dot\a}\, b_k\ ,\label{ghsuper}
\end{eqnarray}
where the index $k$ denotes the level of the ghosts for ghosts. In terms of
these fields, the fully covariant BRST operator can be written as
\begin{equation}
Q={\cal C}_0\Big( {\cal T} + \del {\cal C}_0\, {\cal B}_0 + \sum_{k\ge 1}(
(k+2) \del {\cal C}_k\, {\cal B}_k - (k+1) {\cal C}_k\, \del {\cal B}_k)
\Big) + \sum_{k\ge 0} {\cal C}_{k+1}\, {\cal P}\, {\cal B}_k +
{{\rm ``more"}}\ .\label{fullycobrst}
\end{equation}
The level-$k$ ghosts and antighosts carry an $\a$ index when $k$ is odd and
no index when $k$ is even.   We have suppressed these $\a$ indices, and the
$\a$ index on ${\cal P}_\a$, in the above expression.  The ``more'' term
involves pure-ghost expressions that are needed for the nilpotency of the
BRST operator.   Since the first two terms in the above expression, {\it
i.e.}~the ``BRST'' operator for the original reducible system,
\begin{equation}
Q'={\cal C}_0\, {\cal T} - \del {\cal C}_0\, {\cal C}_0\,{\cal B}_0 \ ,
\label{superbrst}
\end{equation}
is already a nilpotent operator, it implies that the contributions from
higher level ghosts for ghosts are zero.  Indeed, the contributions are zero
level by level, owing to a cancellation between the contributions from the
commuting and anticommuting ghosts for ghosts.

      The complete expression for the BRST operator (\ref{fullycobrst}) is
very complicated, and is not yet known.  Thus a full analysis of its
cohomology is not possible at present.   However, owing to the feature we
discussed above, we can restrict our attention to the physical states with
standard ghost structure in the $k\ge 1$ ghost sectors, {\it i.e.}~the
physical states of the form
\begin{equation}
\ket{\rm phys} = V\ket{0} = V'\ket{\rm gh}_{k\ge1}\ ,\label{statesform}
\end{equation}
where $\ket{\rm gh}_{k\ge1}$ is the standard ghost vacuum for the ghosts for
ghosts, and $V'$ is an operator which involves only the matter and $k=0$
ghosts.  For states of this form, the physical-state condition $Q\ket{\rm
phys}= 0$ involves the following relevant terms in the BRST operator
(\ref{fullycobrst}):
\begin{equation}
Q \sim Q' + \sum_{k\ge0} {\cal C}_{k+1}\, {\cal P}\, {\cal B}_{k}\ .
\label{relevantq}
\end{equation}
Thus for physical states, we must have
\begin{eqnarray}
Q'\, V'\ket{0}&=&0 \ ,\\
(\sum_{k\ge0} {\cal C}_{k+1}\, {\cal P}\, {\cal B}_k)\, V' \ket{\rm
gh}_{k\ge 1} &=&0\ .\label{extracond}
\end{eqnarray}
These states will be non-trivial if the operator $V'$ is
non-trivial with respect to $Q'$.   In the following section, we shall
therefore study the cohomology of the nilpotent charge $Q'$.   Then we shall
discuss the extra condition (\ref{extracond}) on these states to obtain the
physical states of the fully covariant BRST operator (\ref{fullycobrst}).

\section{Cohomology of the reducible system}

      In the previous section, we discussed the fully covariant BRST
quantisation of the reducible system ({\ref{matcur}).   It requires the
introduction of infinitely many ghosts for ghosts.  In this section, we
shall study the cohomology of the BRST operator with the physical states
that are of the form (\ref{statesform}).  As we discussed in the previous
section, for the physical states of this form it is convenient first to
discuss the cohomology of the nilpotent charge $Q'$, and then we shall
examine the extra conditions (\ref{extracond}) arising from the introduction
of the ghosts for ghosts

       For simplicity, we shall discuss the cohomology of the nilpotent BRST
operator $Q'$ in (\ref{superbrst}) in component language.  The matter
current $\cal T$ is defined in (\ref{tsuper}) with components defined in
(\ref{matcur}).  The ghosts and antighosts ${\cal C}_0$, ${\cal B}_0$ are
defined in (\ref{ghsuper}).  We shall from now on suppress the index 0 and
we shall also refer to the nilpotent operator $Q'$ as a ``BRST'' operator.
We introduce anticommuting ghosts $(b,c)$ and $(\beta,\gamma)$ for the bosonic
currents $T$ and $S$, and commuting ghosts $(r^{\dot\alpha},
s_{\dot\alpha})$ for the fermionic currents $G^{\dot\alpha}$, with
$r^{\dot\alpha}(z) s^{\dot\beta}(w)\sim -\epsilon^{\dot\alpha\dot\beta}\,
(z-w)^{-1}$. All anti-ghosts $(b,\beta,r^{\dot \alpha})$ have spin 2, and
all ghosts $(c,\gamma, s_{\dot\alpha})$ have spin $-1$.  Straightforward
computation leads to the following result for the nilpotent operator $Q'$
(\ref{superbrst}) in terms of the component language:
\begin{eqnarray}
Q' &=& Q_0 + Q_1 + Q_2\ ,\\
Q_0 &=& \oint c\Big( -\ft12 \del X^{\alpha\dot\alpha}\, \del X_{\alpha\dot
\alpha} - p_\alpha\, \del\theta^\alpha -b\,\del c -2\beta\,\del\gamma
-\del\beta\, \gamma + 2r^{\dot\alpha}\, \del s_{\dot\alpha} + \del r^{\dot
\alpha}\, s_{\dot\alpha}\Big)\ ,\\
Q_1 &=& \ft12 \oint \gamma\, p_\alpha\, p^\alpha \ ,\\
Q_2 &=& \oint \big(s_{\dot\alpha}\, G^{\dot\alpha} -\beta\, s_{\dot\alpha}
\, \del s^{\dot\alpha} \big)\ .
\label{rsbrst}
\end{eqnarray}

     Under $SL(2,R)_{\rm R}$ transformations, corresponding to the
self-dual Lorentz transformations of the undotted indices, the currents are
all invariant.  Thus the generators of $SL(2,R)_{\rm R}$ are simply given
by
\begin{equation}
J^{\alpha\beta} = \oint \Big(-\ft12 X^{(\alpha}{}_{\dot\alpha}\,\del X^{\beta)
\dot\alpha} + p^{(\alpha}\, \theta^{\beta)} \Big) \ ,\\
\end{equation}
and, for an infinitesimal transformation with (symmetric) parameter
$\omega_{\alpha\beta}$, have the action $\delta \psi^\alpha
=[\omega_{\beta\gamma} J^{\beta\gamma}, \psi^\alpha \}=
\omega^\alpha{}_\beta \, \psi^\beta$ on any undotted index.
The $SL(2,R)_{\rm L}$ transformations, on the other hand, which
correspond to anti-self-dual Lorentz rotations of the dotted indices, rotate
the fermionic currents $G^{\dot\alpha}$, and hence the ghosts $(r^{\dot
\alpha},s_{\dot\alpha})$ must rotate also.  It is quite easy to see that the
generators are given by
\begin{equation}
J^{\dot\alpha\dot\beta} = \oint\Big(\ft12 X^{\alpha(\dot\alpha}\,\del X_\alpha
{}^{\dot\beta)} +  r^{(\dot\alpha}\, s^{\dot\beta)}\Big)\ ,\\
\label{asdrot}
\end{equation}
and they transform dotted indices according to $\delta\psi^{\dot\alpha} =
\omega^{\dot\alpha}{}_{\dot\beta}\, \psi^{\dot\beta}$.
These spacetime Lorentz transformations are symmetries of the
two-dimensional action including ghosts, and they commute with the BRST
charge.

    It is also useful to write down the form of the generators of manifest
spacetime supersymmetry.  They are given by
\begin{eqnarray}
q_\alpha &=& \oint p_\alpha \ ,\\
q^{\dot\alpha} &=& \oint \Big(- \theta_\alpha\, \del X^{\alpha\dot\alpha} -
\gamma\, r^{\dot\alpha} + b\, s^{\dot\alpha} \Big) \ .
\label{rsssusy}
\end{eqnarray}
The somewhat unusual ghost terms in $q^{\dot\alpha}$ are a consequence of
the fact that $r^{\dot\alpha}$ and $s_{\dot\alpha}$ transform under the
anti-self-dual spacetime Lorentz group.  It is straightforward to verify
that these supercharges generate the usual $N=1$
spacetime superalgebra (\ref{n1susyalg}).

     As usual in a theory with fermionic currents, it is appropriate to
bosonise the associated commuting ghosts.  Thus we write
\begin{equation}
r^{\dot\alpha} = \del\xi^{\dot\alpha}\, e^{-\phi_{\dot\alpha}}\ , \qquad
\qquad s_{\dot\alpha}= \eta_{\dot\alpha}\, e^{\phi_{\dot\alpha}}\ ,
\label{boso}
\end{equation}
where $\eta_{\dot\alpha}$ and $\xi^{\dot\alpha}$ are anticommuting fields
with spins 1 and 0 respectively.  The OPEs of the bosonising fields are
$\eta_{\dot\alpha}(z)\xi^{\dot\beta}(w)\sim \delta_{\dot\alpha}^{\dot\beta}
\,(z-w)^{-1}$, and $\phi_{\dot\alpha}(z)\phi_{\dot\beta}(w)\sim
-\delta_{\dot\alpha\dot\beta}\,\log(z-w)$.
Note that the bosonisation breaks the manifest $SL(2,R)_{\rm L}$ covariance,
and that the $\dot\alpha$ index in (\ref{boso}) is not summed.  In view of
this non-covariance, there is no particular advantage in using upper as well
as lower indices on $\phi_{\dot\alpha}$, and we find it more convenient
always to user use lower ones for this purpose.

     The nilpotent operator $Q'$ can be easily re-expressed in terms of the
bosonised fields; the $(r,s)$ terms in $Q_0$ become $\oint
c\Big(-\eta_{\dot\alpha}\, \del \xi^{\dot\alpha} -\ft12(\del\phi_1)^2
-\ft12(\del\phi_2)^2 -\ft32 \del^2\phi_1 -\ft32\del^2\phi_2\Big)$, whilst
$Q_2$ becomes
\begin{eqnarray}
Q_2 &=& \oint \Big(\eta_1\, p_\alpha\, \del X^{\alpha 1}\, e^{\phi_1} +
\eta_2\, p_\alpha\, \del X^{\alpha 2}\, e^{\phi_2} \Big)\nonumber \\
&+& \oint\beta\Big(\eta_1\, \del\eta_2 -\del\eta_1\,\eta_2 - \eta_1\, \eta_2
(\del\phi_1 -\del\phi_2)\Big) e^{\phi_1+\phi_2} \ .
\end{eqnarray}
(It is to be understood that an expression such as $e^{\phi_1+\phi_2}$ really
means $:e^{\phi_1}: \, :e^{\phi_2}:$, which equals $-:e^{\phi_2}: \,
:e^{\phi_1}:$ since both of these exponentials are fermions. Thus we have
$e^{\phi_1+\phi_2}=-e^{\phi_2+\phi_1}$ in this rather elliptical notation.)

    The ghost contributions to the $SL(2,R)_{\rm L}$ Lorentz generators
(\ref{asdrot}) become
\begin{eqnarray}
J_+ &=& r_1\, s_1 = \eta_1\, \del\xi^2\, e^{\phi_1-\phi_2}\ ,\nonumber\\
J_- &=& r_2\, s_2 = \eta_2\, \del\xi^1\, e^{-\phi_1+\phi_2}\ ,
\label{asdghost}\\
J_3 &=& r_{(1}\, s_{2)} = -\ft12(\del\phi_1-\del\phi_2)\ .
\end{eqnarray}
One can easily see that these generate an $SL(2,R)$ Kac-Moody algebra.  The
translation of the supersymmetry charges into bosonised form is obtained by
simple substitution.

     Since the zero modes of the $\xi^{\dot\alpha}$ fields are not included
in the Hilbert space of physical states, there exist BRST non-trivial
picture-changing operators $Z^{\dot\alpha}=\{Q',\xi^{\dot\alpha}\}$ which
can give new BRST non-trivial physical operators when normal ordered with
others.  Explicitly, they take the form
\begin{eqnarray}
Z^1 &=& c\, \del\xi^1 - p_\alpha\, \del X^{\alpha 1}\, e^{\phi_1} -
 \Big(2\beta\, \del\eta_2 + \del\beta\,\eta_2 +2\beta\, \eta_2\,\del\phi_2
\Big) e^{\phi_1+\phi_2}\ ,\\
Z^2 &=& c\, \del\xi^2 - p_\alpha\, \del X^{\alpha 2} \, e^{\phi_2} -
 \Big(2\beta\, \del\eta_1 + \del\beta\,\eta_1 +2\beta\, \eta_1\,\del\phi_1
\Big) e^{\phi_1+\phi_2}\ ,
\label{pic}
\end{eqnarray}
Like the case of the irreducible system, these operators have no inverse.

\subsection{Physical states of $Q'$}

\subsubsection{Preliminaries}

     In this subsection, we shall discuss the cohomology of the nilpotent
operator $Q'$.  Owing to the rather  unusual feature in this theory that
some of the ghosts carry target spacetime  spinor indices, the notion of the
standard ghost vacuum requires some  modification.  We begin by noting that
the non-vanishing correlation function that defines the meaning of
conjugation is given by
\begin{equation}
\Big\langle \del^2c\,\del c\, c\,\del^2\gamma\, \del\gamma\, \gamma\,
e^{-3\phi_1-3\phi_2}\, \theta^2\, \Big\rangle \ne 0\ ,
\label{prod}
\end{equation}
where $\theta^2\equiv\theta^\alpha\, \theta_\alpha$.
In terms of the bosonised form of the commuting
ghosts, the usual operator $e^{-\phi_1-\phi_2}$ appearing in the definition
of the ghost vacuum can be generalised to an operator $W_{\dot\alpha_1\cdots
\dot\alpha_{2s}}$, totally symmetric in its indices, whose component with
$(s+m)$ indices taking the value $\dot1$ and $(s-m)$ taking the value $\dot2$
is given by
\begin{equation}
W_{\dot1\cdots \dot1 \dot2\cdots \dot2} =
\lambda(s,m)\,\del^{s+m-1}\eta_1 \cdots
\del\eta_1 \, \eta_1\,  \del^{s-m-1}\eta_2\cdots
\del\eta_2 \, \eta_2\, e^{(s+m-1)\phi_1 +(s-m-1)\phi_2}\ .
\label{Wdef}
\end{equation}
The normalisation constants $\lambda(s,m)$ are given by
\begin{equation}
\lambda(s,m) = \prod_{p=1}^{s+m-1}\ \prod_{q=1}^{s-m-1} {1\over p!\, q!}\ ,
\end{equation}
where any product over an empty range is defined to be 1. $W$ in (\ref{Wdef})
has $(s+m)$ factors involving $\eta_1$, and $(s-m)$ factors involving
$\eta_2$, with $-s\le m\le s$.  It is the $J_3=m$ state in the
$(2s+1)$-dimensional spin-$s$ representation of $SL(2,R)_{\rm L}$.  The
operator $W_{\dot1\cdots \dot1}$ corresponds to the highest-weight state in the
representation, satisfying $J_+\, W_{\dot1\cdots \dot1}=0$,
with the remaining $2s$
states being obtained by acting repeatedly with $J_-$, each application of
which turns a further ``$\dot1$'' index into a ``$\dot2$'',
until the lowest-weight
state $W_{\dot2\cdots \dot2}$ is obtained.  Note, incidentally,
that the form of the
states given in (\ref{Wdef}) becomes rather simple if one bosonises the
$(\eta,\xi)$ fields.

     We may also define a ``conjugate'' operator $\widetilde
W^{\dot\alpha_1\cdots\dot\alpha_{2s}}$, again totally symmetric in its
indices, by
\begin{equation}
\widetilde W^{\dot1\cdots \dot1 \dot2\cdots \dot2} =
\tilde\lambda(s,m)\,\del^{s+m}\xi^1\cdots
\del^2\xi^1 \, \del\xi^1\,  \del^{s-m}\xi^2\cdots
\del^2\xi^2 \, \del\xi^2 e^{-(s+m+2)\phi_1 -(s-m+2)\phi_2}\ ,
\end{equation}
with
\begin{equation}
\tilde\lambda(s,m)= \prod_{p=1}^{s+m}\ \prod_{q=1}^{s-m} {1\over p!\, q!}\ .
\end{equation}
Thus the usual ghost vacuum operator $e^{-\phi_1-\phi_2}$ and its
``conjugate'' $e^{-2\phi_1-2\phi_2}$ correspond to the $s=0$ cases $W$ and
$\widetilde W$ respectively.  All the operators
$W_{\dot\alpha_1\cdots\dot\alpha_{2s}}$ and $\widetilde
W^{\dot\alpha_1\cdots\dot\alpha_{2s}}$ have worldsheet conformal spin 2, and
they all have the property of defining vacuum states that are annihilated by
the positive Laurent modes of $r^{\dot\alpha}$ and $s_{\dot\alpha}$, but not
by the negative modes.

    These operators have simple properties when acted on by $Q'$.
The relevant facts can be summarised in the following lemmas:
\begin{eqnarray}
Q_2\, W_{\dot\alpha_1\cdots \dot\alpha_{2s}}\, \theta^\alpha\,e^{ip\cdot X} &=&
i p^{\alpha\dot\alpha_{2s+1}}\,
W_{\dot\alpha_1\cdots \dot\alpha_{2s+1}}\,
e^{ip\cdot X}\ ,\\
Q_2\, \widetilde W^{\dot\alpha_1\cdots \dot\alpha_{2s}}\, \theta^\alpha\,
e^{ip\cdot X} &=&
i p^{\alpha(\dot\alpha_1}\, \widetilde W^{\dot\alpha_2\cdots
\dot\alpha_{2s})}\,
e^{ip\cdot X}\ .
\label{lemmas}
\end{eqnarray}
A factor of $\gamma$ or $\del\gamma\gamma$ may be included on both sides of
the equation in either formula.  Note that in (\ref{lemmas}), the right-hand
side is defined to be zero if $s=0$.   It is worth remarking that
 we have recovered the manifest
covariance under $SL(2,R)_{\rm L}$ in the expressions for the
$W^{\dot\alpha_1\cdots\dot\alpha_{2s}}$ and $\widetilde W^{\dot\alpha_1
\cdots\dot\alpha_{2s}}$, even though it was broken by the bosonisation of the
ghosts.

\subsubsection{Physical states}

     Let us first consider massless operators in the physical spectrum.
There are four types of massless operators that can be built from $W$
operators, namely
\begin{eqnarray}
U&=&h_{\alpha\dot\alpha_1\cdots \dot\alpha_{2s}}\, c\, \gamma\,
W^{\dot\alpha_1\cdots \dot\alpha_{2s}}\, \theta^\alpha\,
e^{ip\cdot X}\ ,\nonumber\\
V&=&g_{\alpha\dot\alpha_1\cdots \dot\alpha_{2s}}\, c\, \del\gamma\,\gamma\,
W^{\dot\alpha_1\cdots \dot\alpha_{2s}}\, \theta^\alpha\,
e^{ip\cdot X}\ , \nonumber\\
\Psi&=&h_{\dot\alpha_1\cdots \dot\alpha_{2s}}\, c\, \gamma\,
W^{\dot\alpha_1\cdots \dot\alpha_{2s}}\,
e^{ip\cdot X}\ , \label{Umult}\\
\Phi&=&g_{\dot\alpha_1\cdots \dot\alpha_{2s}}\, c\,\del\gamma\, \gamma\,
W^{\dot\alpha_1\cdots \dot\alpha_{2s}}\,
e^{ip\cdot X}\ , \nonumber
\end{eqnarray}
and there are four types of physical operators that are  associated with the
operator $\widetilde W$
\begin{eqnarray}
\widetilde U&=&\tilde h_{\alpha\dot\alpha_1\cdots \dot\alpha_{2s}}\, c\,
\gamma\,  \widetilde W^{\dot\alpha_1\cdots \dot\alpha_{2s}}\, \theta^\alpha\,
e^{ip\cdot X}\ ,\nonumber\\
\widetilde V&=&\tilde g_{\alpha\dot\alpha_1\cdots \dot\alpha_{2s}}\, c\,
\del\gamma\,\gamma\, \widetilde W^{\dot\alpha_1\cdots \dot\alpha_{2s}}\,
\theta^\alpha\,  e^{ip\cdot X}\ , \nonumber\\
\widetilde\Psi&=&\tilde h_{\dot\alpha_1\cdots \dot\alpha_{2s}}\, c\, \gamma\,
\widetilde W^{\dot\alpha_1\cdots \dot\alpha_{2s}}\,\theta^2\,
e^{ip\cdot X}\ , \label{UUmult}\\
\widetilde\Phi&=&\tilde g_{\dot\alpha_1\cdots \dot\alpha_{2s}}\,
c\,\del\gamma\, \gamma\,  \widetilde W^{\dot\alpha_1\cdots
\dot\alpha_{2s}}\,\theta^2\,   e^{ip\cdot X}\ .\nonumber
\end{eqnarray}
These operators are the conjugates of the physical operators (\ref{Umult})
with $\del c\, c \longrightarrow c$.  We shall thus only discuss the
physical-state condition of the states in (\ref{Umult})

     We find that $U$ itself is annihilated by $Q'$ provided that the
following conditions hold:
\begin{equation}
p^{\alpha\dot\alpha}\, p_{\alpha\dot\alpha}=0\ ,\quad \qquad
h^{\alpha(\dot\alpha_1\cdots\dot\alpha_{2s}}\, p_\alpha{}^{\dot\alpha)} =0\ .
\label{eqh}
\end{equation}
The first of these is just the mass-shell condition for massless states.
Having ensured that $U$ is annihilated by $Q'$, we must also check to see
whether it is BRST non-trivial.  One way to do this is by constructing
conjugate operators that have a non-vanishing inner product with $U$, as
defined by (\ref{prod}).  If the inner-product is non-vanishing for
conjugate operators that are annihilated by $Q'$, then $U$ is BRST
non-trivial.  Operators $U^\dagger$ conjugate to $U$ have the form
\begin{equation}
U^\dagger = f_{\alpha\dot\alpha_1\cdots\dot\alpha_{2s}}\, \del c\, c\,
\del\gamma\, \gamma\, \widetilde W^{\dot\alpha_1\cdots\dot\alpha_{2s}}\,
\theta^\alpha\, e^{ip\cdot X}\ ,
\end{equation}
which is annihilated by $Q'$ if
\begin{equation}
p^{\alpha\dot\alpha_1}\, f_{\dot\alpha_1\cdots\dot\alpha_{2s}}=0\ .
\label{eqf}
\end{equation}
It is convenient to choose a particular momentum frame in order to analyse
the true physical degrees of freedom that are implied by these kinematical
conditions.  The null momentum vector $p^\mu$ may, without loss of
generality, be chosen to be $p^\mu=(1,0,0,1)$.  From (\ref{vdw}), this
implies that all components of $p^{\alpha\dot\alpha}$ are zero except for
$p^{1\dot1}=\sqrt2$.  In this frame, the solutions to (\ref{eqh}) and
(\ref{eqf}) are
\begin{equation}
h_{1\dot\alpha_1\cdots\dot\alpha_{2s}}=0\ ,\qquad
f_{1 \dot1\dot\alpha_2\cdots\dot\alpha_{2s}} =0 \ .
\label{solhf}
\end{equation}
The inner product has the form $\langle U^\dagger\, U\rangle= f^{\alpha
\dot\alpha_1\cdots\dot\alpha_{2s}}\, h_{\alpha \dot\alpha_1\cdots\dot
\alpha_{2s}}=
f^{2\dot1\cdots \dot1}\, h_{2\dot1\cdots \dot1}$.  Thus there is just one
physical degree of freedom described by $U$, corresponding to the
polarisation spinor component $h_{2\dot1\cdots \dot1}$.  The other
non-vanishing components of $h_{\alpha\dot\alpha_1\cdots\dot\alpha_{2s}}$
allowed by
(\ref{solhf}) correspond to BRST trivial states, and can be expressed back in
covariant language as pure-gauge states with
\begin{equation}
h_{\alpha\dot\alpha_1\cdots\dot\alpha_{2s}}=p_{\alpha(\dot\alpha_1}\,
\Lambda_{\dot\alpha_2\cdots\dot\alpha_{2s})}\ ,\label{hguage}
\end{equation}
where $\Lambda_{\dot\alpha_2\cdots\dot\alpha_{2s}}$ is arbitrary.
We note also, for
future reference, that the equation of motion for
$h^{\alpha\dot\alpha_1\cdots\dot\alpha_{2s}}$ in (\ref{eqh}) is equivalent to
\begin{equation}
h^{\alpha\dot\alpha_1\cdots\dot\alpha_{2s}}\, p_\alpha{}^{\dot\alpha} =0\ .
\label{nonsym}
\end{equation}

     The operator $\Psi$ in (\ref{Umult}) is annihilated by $Q'$ provided
just that the mass-shell condition $p^{\alpha\dot\alpha}\,
p_{\alpha\dot\alpha}=0$ is satisfied.  To see the physical degrees of
freedom, we again consider conjugate operators $\Psi^\dagger$, which have
the form $\Psi^\dagger=f_{\dot\alpha_1\cdots\dot\alpha_{2s}}\,
\del c\, c\, \del
\gamma\, \gamma\, \widetilde W^{\dot\alpha_1\cdots\dot\alpha_{2s}}\,
\theta^2\,
e^{ip\cdot X}$.  This is annihilated by $Q'$ provided that $p^{\alpha\dot
\alpha_1}\, f_{\dot\alpha_1\cdots\dot\alpha_{2s}}=0$. In the special momentum
frame, the solution is $f_{\dot1\dot\alpha_2\cdots\dot\alpha_{2s}}=0$. Thus the
inner product is proportional to $h^{\dot2\cdots \dot 2}\,
f_{\dot2\cdots \dot 2}$, so only the one component  $h^{\dot2\cdots \dot 2}$
describes a true physical degree of freedom.  The unphysical BRST-trivial
components correspond to polarisation spinors of the pure-gauge form
\begin{equation}
h^{\dot\alpha_1\cdots\dot\alpha_{2s}}=p^{\alpha(\dot\alpha_1}\,
\Lambda_\alpha{}^{\dot\alpha_2\cdots\dot\alpha_{2s})}\ .\label{gguage}
\end{equation}
The analysis of the operators $V$ and $\Phi$ in (\ref{Umult}) is precisely
the same as the above analyses for $U$ and $\Psi$ respectively.

    So far, we have concentrated on massless states in the physical
spectrum.  There are also massive physical states, an example
being $c\, e^{-\phi_1-\phi_2}\, e^{ip\cdot X}$ with $p^{\alpha\dot\alpha}\,
p_{\alpha\dot\alpha}=-2$, implying (mass)$^2 = 2$.  Further examples are
\begin{equation}
V= c\, \del^{2n}\beta\cdots \del\beta\, \beta\, (\del^n p)^2\cdots (\del
p)^2\, p^2\, e^{n(\phi_1+\phi_2)} \, e^{ip\cdot X}\ ,\label{massivefull}
\end{equation}
where $p^2=p^\alpha\, p_\alpha$, {\it etc}.  These spacetime scalar states
are physical for arbitrary integer $n$, with (mass)$^2 = 2(n+1)(2n+3)$.

      Another class of physical states in the theory is associated with
infinite-dimensional representations of $SL(2,R)_{\rm L}$. Consider, for
example, the operators
\begin{eqnarray}
X&=&c\, e^{-\phi_1}\, e^{ip\cdot X}\ ,\nonumber\\
Y&=&c\, \del\gamma\, \gamma\, \theta^2\,e^{-2\phi_1-\phi_2}\, e^{ip\cdot X}\ .
\label{VB21}
\end{eqnarray}
This is annihilated by the BRST operator provided that the mass-shell
condition $p^{\alpha\dot\alpha} p_{\alpha\dot\alpha}=0$ is satisfied,
together with the transversality condition $p^{\alpha \dot2}=0$.  This
condition is not covariant with respect to $SL(2,R)_{\rm L}$, suggesting
that further terms should be added in order to construct a
fully-covariant physical operator.  This is analogous to viewing a physical
operator built using $W_{\dot\alpha_1\cdots\dot\alpha_{2s}}$ as consisting
of the term involving $W_{\dot1\cdots \dot1}$ plus the remaining $2s$ terms
obtained by acting repeatedly on this highest-weight state with $J_-$.
Thus, noting that $e^{-2\phi_1-\phi_2}$ is a highest-weight state, $J_+\,
e^{-2\phi_1-\phi_2} =0$, we may replace (\ref{VB21}) by the $SL(2,R)_{\rm
L}$ covariant operators
\begin{eqnarray}
X&=& \sum_{n\ge 0} h_n\, c\, \Big((J_{-})^n\, e^{-\phi_1})\, e^{ip\cdot X}
\ ,\nonumber \\
Y&=&\sum_{n\ge 0} h_n\, c\, \del\gamma\,\gamma\, \theta^2\, \Big((J_-)^n
e^{-2\phi_1-\phi_2}\Big)\, e^{ip\cdot X}\ .\label{VB21full}
\end{eqnarray}
One can easily see from the form of the generator $J_-$ in (\ref{asdghost})
that in this case the process of repeated application of $J_-$ will never
terminate, and the sum over $n$ will be an infinite one, corresponding to an
infinite-dimensional representation of $SL(2,R)_{\rm L}$.  The
physical-state conditions will now give a transversality condition on the
components $h_n$ of the polarisation tensor, rather than the non-covariant
condition $p^{\alpha\dot 2}=0$ that resulted when only the $n=0$ term was
included.  The occurrence of infinite-dimensional representations of
$SL(2,R)_{\rm L}$ is an undesirable feature of the theory.  We shall see
later, the extra conditions (\ref{extracond}) will eliminate these types of
states.

\subsection{Cohomology of the reducible system}

    In the previous subsections, we obtained some examples of physical
states for the nilpotent operator $Q'$.  Not all of them, however, can  give
rise to physical states of the fully covariant BRST operator $Q$ according
to the prescription given in (\ref{statesform}).  In this subsection, we
shall examine the extra conditions (\ref{extracond}) on these states in
order to determine which states will be eliminated.

    For the states of the form (\ref{statesform}), the matter term ${\cal
P}_\a$ only acts on the operator $V'$, and always gives rise to a first
order pole in the operator product expansion with the massless physical
operators we discussed in the previous subsections.  The ghost terms ${\cal
C}_{k+1}\, {\cal B}_k$ for $k\ge 1$ in the extra conditions
(\ref{extracond}) will only act on the ghost for ghost operator, which has
standard ghost structure, and will give rise to a first order zero in the
operator product expansion.  Thus these terms in ({\ref{extracond}) will
annihilate the states.  However, when $k=0$, the ghost term ${\cal C}_1\,
{\cal B}_0$ will give rise to a zeroth order pole in the operator product
expansion with the $V'$ given by either (\ref{VB21}) or (\ref{VB21full}) or
their conjugates, whilst it gives a first order zero with the $V'$
corresponding to any of the other massless operators we discussed
previously.  Thus the extra conditions arising from the ghosts for ghosts
eliminate the states (\ref{statesform}) with $V'$ given by the massless
operators that are infinite-dimensional representations of $SL(2,R)_{\rm
L}$, given by (\ref{VB21}) and (\ref{VB21full}).   The massive operators
(\ref{massivefull}) will also be eliminated by the extra conditions since
the ${\cal P}_\a$ term will produce higher order poles in the operator
product expansions.   However this does not necessarily imply that there are
no massive operators in the the physical spectrum of the reducible system
since we have only considered physical states with standard ghost structure
for the ghosts for ghosts.

       In order to compare the physical spectra of the nilpotent operator
$Q'$ of the reducible system and the BRST operator $Q$ of the irreducible
system,  it is instructive to discuss the interactions. Interactions amongst
the physical states that we have found so far are not easy to come by.  One
example that can occur is a three-point interaction $\langle U U \Psi
\rangle$ between two fermions and a boson in the scalar supermultiplet
corresponding to the $s=0$ operators, as given in (\ref{Umult}). The
four-point amplitudes involving only the above operators are zero.  This is
precisely the same result we obtained for the irreducible system. For higher
values of $s$, interactions necessarily involve tilded physical states and
picture-changing operators.  However all the tilded physical states have
vanishing normal-ordered products with the picture-changing operators, and
thus there are no interactions among these states.  Thus all the physical
operators with higher values of $s$ decouple from the theory.

      The physical operators under $Q'$ that carry infinite-dimensional
representations of $SL(2,R)_{\rm L}$, however can have interactions with the
$s=0$ spinor operators.  For example, we can build a non-vanishing
four-point amplitude $\langle U\, V\, X\, X\rangle$ without the need for the
picture changing, where $U,V$ are given by (\ref{Umult}) with $s=0$, and $X$
is given by (\ref{VB21full}).   This would be inconsistent with the results
of the irreducible system, where no non-vanishing four-point amplitude can
be built from purely massless operators.  However, this apparent discrepancy
is overcome by the introduction of the ghosts for ghosts, since as we
discussed above, these operators will be eliminated by the extra conditions
in ({\ref{extracond}).

\section{Discussion and conclusions}

     In this paper, we have constructed a superstring theory in
four-dimensional spacetime with $(2,2)$ signature, using the Berkovits'
approach of augmenting the spacetime supercoordinates by the conjugate momenta
for the fermionic variables \cite{Berk}.  The form of the theory,
and its local worldsheet symmetries, was motivated by Siegel's proposal
\cite{S1} for a set of constraints that could give rise to self-dual super
Yang-Mills theory or supergravity in $2+2$ dimensions.  In the theory that we
have considered, $N=1$ spacetime supersymmetry is manifest in the
formulation, as is the right-handed $SL(2,R)$ factor of the $SO(2,2)\equiv
SL(2,R)_{\rm L} \times SL(2,R)_{\rm R}$ Lorentz group.

     The constraints that we have used are a subset of Siegel's constraints
\cite{S1} that form a closed algebra under commutation.  They are manifestly
$SO(2,2)$ covariant; however, they suffer from the fact that they are
reducible, with an infinite order of reducibility.  This implies that the
fully covariant BRST treatment requires infinitely many ghosts for
ghosts.    Nevertheless, we can construct the nilpotent operator for the
reducible system.   We studied the cohomology of this nilpotent operator.
It gives rise to a theory with an infinite number of massless states with
arbitrary spin.  This feature can be attributed to the fact that the
fermionic constraint carries an $SL(2,R)_{\rm L}$ spinor index, leading to
the existence of ghost vacua with arbitrary spin $s$ under $SL(2,R)_{\rm L}$.
However these higher-spin massless states are decoupled from the $s=0$
states.

       One way to overcome the reducibility problem is by choosing an
irreducible subset of constraints,  at the cost of breaking the spacetime
Lorentz covariance.   Since the irreducible constraints also have critical
central charge in our case, we were able to construct the BRST operator and
study its cohomology.   We showed that the theory also maintains the
manifest $N=1$ supersymmetry.  The physical spectrum with standard ghost
structure includes a massless bosonic operator and its superpartner.
Their interactions comprise only a three-point amplitude with one insertion
of the boson and two insertions of the fermion.   This pattern of
interactions is precisely the same as that of the $s=0$ massless states of
the reducible system.

\centerline{\bf Acknowledgements}
\bigskip

    H.L.\ and C.N.P.\ are grateful to SISSA, Trieste, and E.S. is grateful
to the ICTP, Trieste, for hospitality during the course of this work.  We
are grateful to the referee for pointing out the reducibility of the
constraints in the the realisation (\ref{matcur}).

\end{document}